\documentclass[conference]{IEEEtran}
\IEEEoverridecommandlockouts

\usepackage{cite}
\usepackage{amsmath,amssymb,amsfonts}
\usepackage{algorithmic}
\usepackage{graphicx}
\usepackage{textcomp}
\usepackage{xcolor}
\usepackage[hidelinks]{hyperref}
\usepackage{booktabs}

\def\BibTeX{{\rm B\kern-.05em{\sc i\kern-.025em b}\kern-.08em
    T\kern-.1667em\lower.7ex\hbox{E}\kern-.125emX}}
\begin{document}

\title{%
Modelling Pedestrian Behaviour in Autonomous Vehicle Encounters Using Naturalistic Dataset\\[0.3em]
\large\textbf{IEEE Intelligent Vehicles Symposium (IV), June 22--25, 2026, Detroit, MI, United States}\\[-0.6em]
\thanks{This research is funded by Canada Research Chair in Disruptive Transportation Technologies and Services (CRC-2022-00480).}
}

\author{\IEEEauthorblockN{Rulla Al-Haideri and Bilal Farooq}
\IEEEauthorblockA{\textit{Laboratory of Innovations in Transportation (LiTrans)} \\
\textit{Toronto Metropolitan University}, \textit{Canada} \\
\{rullaalhaideri, bilalfarooq\}@torontomu.ca}
}

\maketitle

\begin{abstract}
Understanding how pedestrians adjust their movement when interacting with autonomous vehicles (AVs) is essential for improving safety in mixed traffic. This study examines micro-level pedestrian behaviour during midblock encounters in the NuScenes dataset using a hybrid discrete choice--machine learning framework based on the Residual Logit (ResLogit) model. 
The model incorporates temporal, spatial, kinematic, and perceptual indicators. These include relative speed, visual looming, remaining distance, and directional collision risk proximity (CRP) measures. Results suggest that some of these variables may meaningfully influence movement adjustments, although predictive performance remains moderate. Marginal effects and elasticities indicate strong directional asymmetries in risk perception, with frontal and rear CRP showing opposite influences. The remaining distance exhibits a possible mid-crossing threshold. Relative speed cues appear to have a comparatively less effect. These patterns may reflect multiple behavioural tendencies driven by both risk perception and movement efficiency. 
\end{abstract}

\begin{IEEEkeywords}
Reslogit, pedestrians, autonomous vehicles, hybrid behavioural models, visual looming, collision risk proximity
\end{IEEEkeywords}

\section{Introduction}\label{intro}
Pedestrian safety remains a critical concern in mixed traffic where human driven vehicles and autonomous vehicles (AVs) coexist. Among all road interactions, those involving pedestrians present the highest level of risk. Unsignalized crosswalks and midblock crossings are particularly critical, as they expose vulnerable users directly to approaching vehicles. Pedestrian crossing behaviour involves complex perceptual and cognitive processes. These include estimating vehicle trajectories, assessing risk, and adapting to contextual cues such as vehicle type and speed. Understanding these processes is essential for developing AVs that can interact safely and predictably with pedestrians. A large body of research has examined pedestrian–vehicle interactions using field data, video trajectories, and virtual reality experiments. Earlier studies analysed how pedestrians decide to cross based on speed, distance, and gap acceptance. Statistical models such as logistic and multinomial logit have been widely used to describe these behaviours and evaluate the influence of contextual factors \cite{zhou2013multinomial,asaithambi2016pedestrian,shaaban2021midblock}. Although these models provided interpretability, they were limited in capturing nonlinear effects and individual variability in decision making.
Discrete choice models (DCMs) are behavioural tools most commonly formulated under random utility maximization. Individuals are assumed to select the alternative that provides the highest perceived utility. These models have long offered an interpretable way to represent individual preferences and context specific trade offs at a detailed aggregation level. Their ability to quantify how observable attributes and latent factors shape decisions has made them fundamental in transportation research. However, conventional formulations typically rely on linear utility specifications. This limits their ability to capture nonlinear patterns and dynamic effects often present in human decision making. 

The Residual Logit (ResLogit) model proposed by Wong and Farooq \cite{wong2021reslogit} was developed to address several limitations of traditional discrete choice specifications. As a hybrid and interpretable framework, ResLogit embeds residual neural network components within the random utility maximization structure, allowing nonlinear transformations of inputs while retaining behavioural meaning. This enables the model to capture systematic patterns in unobserved heterogeneity that standard linear-in-parameters forms may miss. Building on this foundation, Kamal and Farooq \cite{kamal2024ordinal} extended the approach to ordered outcomes through the Ordinal-ResLogit architecture, which offered improved flexibility for modelling perceptions such as waiting time and safety. Previous work by Kalatian and Farooq \cite{kalatian2019deepwait} introduced DeepWait, a deep survival model for estimating pedestrian waiting time before initiating a crossing. More recently, Kamal and Farooq \cite{kamal2024ordinal} applied the Ordinal-ResLogit to stated preference evaluations of waiting time under different traffic and communication conditions. These studies provide important insights into pre-movement decisions, but they do not capture the fine grained, moment-to-moment adjustments that occur once the pedestrian is already in the crossing. 
To the best of our knowledge, no prior study has applied the family of ResLogit models to examine the micro-level movement behaviour of pedestrians in a naturalistic dataset as they navigate a crossing. Using this framework, therefore, offers an opportunity to complement existing work by analyzing how pedestrians adjust their speed in real time in response to an approaching AV. This may enrich current understanding of pedestrian–AV interactions by capturing behavioural patterns at a level of temporal detail not previously examined.

The present study extends this line of research by analyzing the full pedestrian crossing at a fine temporal resolution. Instead of predicting a single waiting duration, we model the sequence of instantaneous movement choices that pedestrians make throughout the crossing. This framework captures multiple layers of decision hierarchies, including moment-to-moment adjustments in deceleration and acceleration in response to approaching AVs. The analysis incorporates detailed spatial, temporal, and perceptual indicators extracted at each time step to examine how pedestrians regulate their motion under different interaction geometries.

The primary objective of this study is to understand how pedestrians adapt their micro-level speed change behaviour in response to approaching AVs. The analysis focuses on moment-to-moment adjustments that pedestrians make while crossing in front of oncoming AVs. The following research questions guide the study:
\begin{itemize}
    \item How do pedestrians modulate their speed when interacting with an AV at midblock crossings?
    \item Which spatial, temporal, and perceptual variables mostly influence these instantaneous movement decisions?
    \item Do pedestrians exhibit distinct directional or temporal sensitivities in perceived risk, particularly between frontal and rear collision exposure?
\end{itemize}

\section{Dataset and Methods}

We analyzed pedestrian–AV interactions using the NuScenes dataset \cite{caesar2020nuscenes}, which provides multimodal AV data collected in Singapore and Boston. The dataset includes high resolution trajectories of vehicles, pedestrians, and other road users. For the purpose of this study, we extracted all adult pedestrians who completed a full crossing in front of an AV. Crossings were identified using two criteria: geometric intersection between the pedestrian trajectory and the AV path, and cases where the pedestrian completed a lateral movement across the AV's forward path. Following data cleaning and filtering, 137 unique pedestrians met the selection criteria. These trajectories corresponded to 1,875 valid decision time steps, depending on each pedestrian’s start and end positions within the scene. We resampled all trajectories to a 1s interval and computed spatial, temporal, and kinematic variables at each time instant of the pedestrian. The sample was small, and the 1s resampling may smooth fast reactions and related indicators.

To characterize the interaction between the pedestrian and the approaching AV, we derived a set of perceptual and conflict based indicators. Visual looming was computed from the change in the AV’s apparent angular size as projected in the pedestrian’s field of view. The relative looming rate, defined as the percentage change in angular size per second $\big(100\frac{\dot{\theta}}{\theta}\big)$, expresses how quickly the approaching AV appears to grow in the pedestrian’s eye and therefore implies the urgency of its approach. A small value of this metric indicates that the AV’s apparent size is changing slowly. This means that the vehicle is either far away or approaching at a gentle rate. We hypothesize that under these conditions, the pedestrian perceives little visual pressure to adjust their movement.
A large value, on the other hand, means the AV’s projected size is increasing rapidly. This occurs when the vehicle is closing in quickly, creating a strong sense of urgency and signalling a more immediate potential conflict from the pedestrian’s perspective.

Relative speed is defined as the speed of the AV relative to the pedestrian. It is computed as the magnitude of the difference between the AV’s velocity vector and the pedestrian’s velocity vector at each time step.
A small relative speed indicates that the AV is closing in slowly and the interaction is less urgent. A large relative speed reflects a rapidly decreasing separation between the pedestrian and the AV and implies a higher level of approach urgency.

Closing Time-to-Collision (CTTC) represents the time remaining until the pedestrian and the approaching AV would collide if both maintained their current headings and speeds \cite{alhaideri2025cttc}. CTTC is calculated along the pedestrian’s line of sight to the AV and takes positive values only when the vehicle is visually closing in. A small CTTC indicates that, from the pedestrian’s eye, the AV appears to be approaching imminently and the interaction feels more urgent. A large CTTC corresponds to a visually slower or more distant approach. Because CTTC decreases as risk increases, its inverse forms the basis of the collision intensity measures.

\begin{figure}[!ht]
    \centering
    \includegraphics[clip=false, width=0.5\textwidth, keepaspectratio]{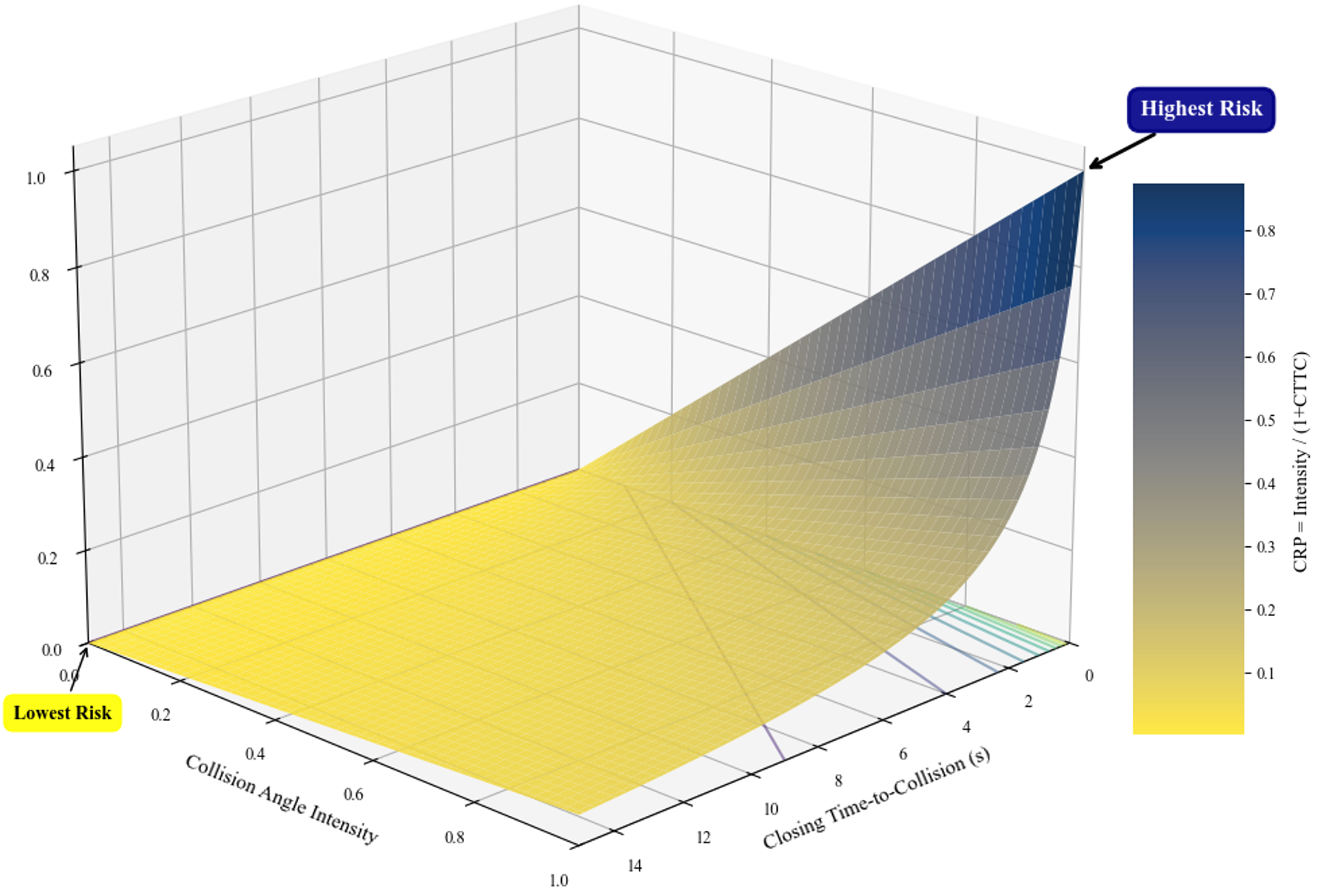}
    \caption{Directional visual risk surface as a function of collision angle intensity and CTTC, computed as $\text{Intensity}/(1+\text{CTTC})$.}
    \label{fig:directional_risk_surface}
\end{figure}

The collision angle is obtained as the azimuth between the pedestrian’s velocity vector and the vector pointing from the pedestrian to the AV. This angle indicates where the AV lies relative to the pedestrian’s walking direction. We transform this angle into two directional intensities. The frontal collision angle intensity (CAI) quantifies how directly the AV appears in the pedestrian’s forward field of view. It is obtained by taking the cosine of the collision angle when the AV lies within the frontal sector and zero otherwise. A rear sector intensity is defined analogously for AVs approaching from behind. Finally, CTTC and the directional CAI are combined into a single spatio-temporal measure, referred to as the Collision Risk Proximity (CRP) indicator. The CRP is defined as the CAI divided by $(1+\text{CTTC})$. This formulation increases the perceived risk when the AV lies in a direction of high CAI, whether within the pedestrian’s forward field or the rear field, and when the visual closing time is short (small CTTC). 
We also capture short term temporal dynamics using two mechanisms. First, the recent change in relative speed was defined as a 3-sample moving average of the one step difference in relative speed along each pedestrian trajectory. Second, we included 3 step lagged versions of the front and rear CRP terms, defined as the CTTC weighted intensities evaluated three seconds earlier, to represent delayed effects of previous exposure rather than relying solely on instantaneous risk.
Figure~\ref{fig:directional_risk_surface} illustrates this relationship and shows that directional visual risk increases sharply when CAI is high and CTTC is small.

\vspace{-1 mm}
\setlength{\textfloatsep}{8pt plus 1pt minus 2pt}

\begin{figure}[!ht]
    \centering
    \includegraphics[clip=false, width=0.5\textwidth, keepaspectratio]{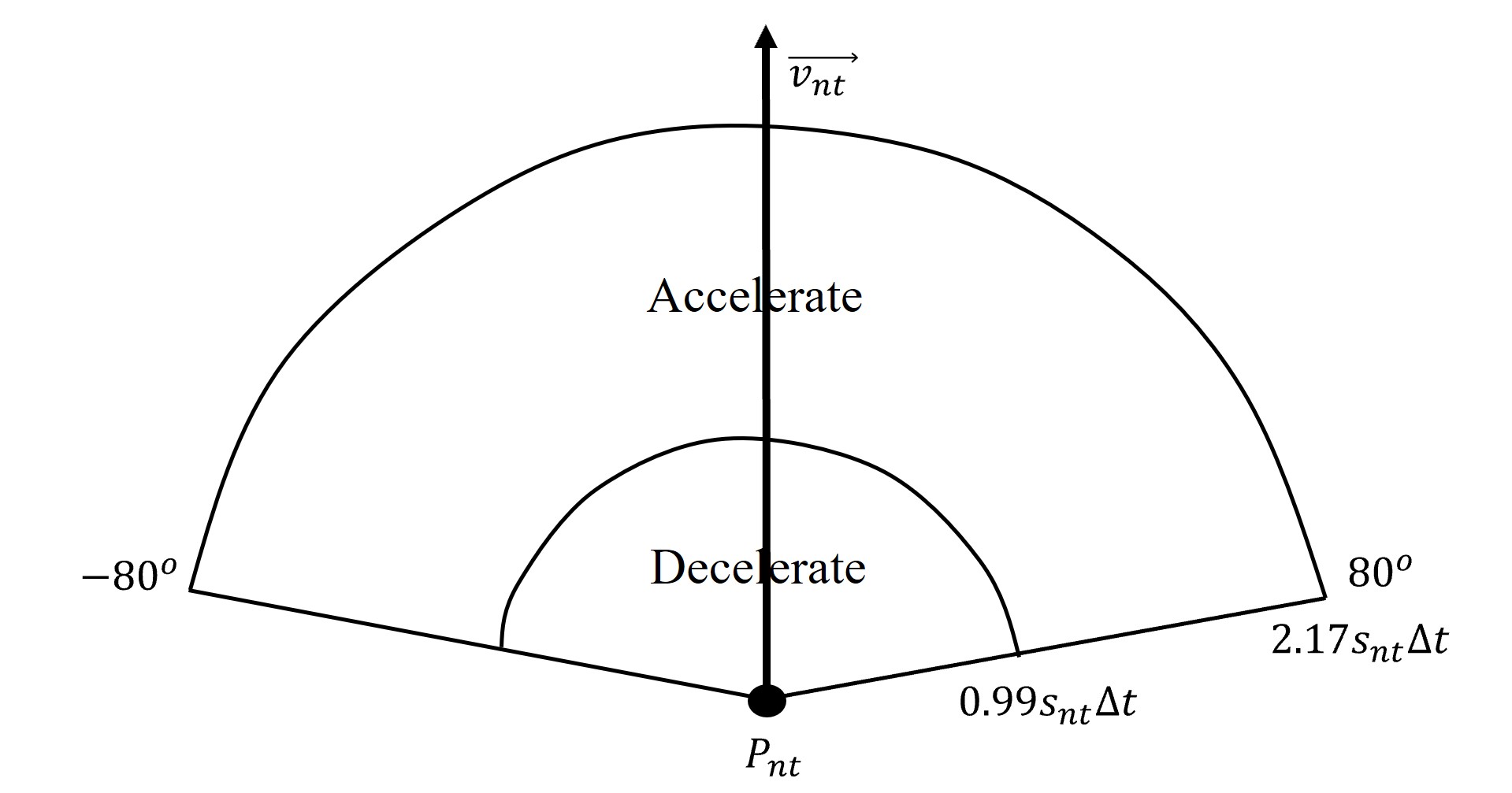}
    \caption{Spatial choice grid structure in front of each pedestrian position $P_{nt}$.}
    \label{fig:grid_def}
\end{figure}

\begin{figure}[!ht]
    \centering
    \includegraphics[clip=false, width=0.5\textwidth, keepaspectratio]{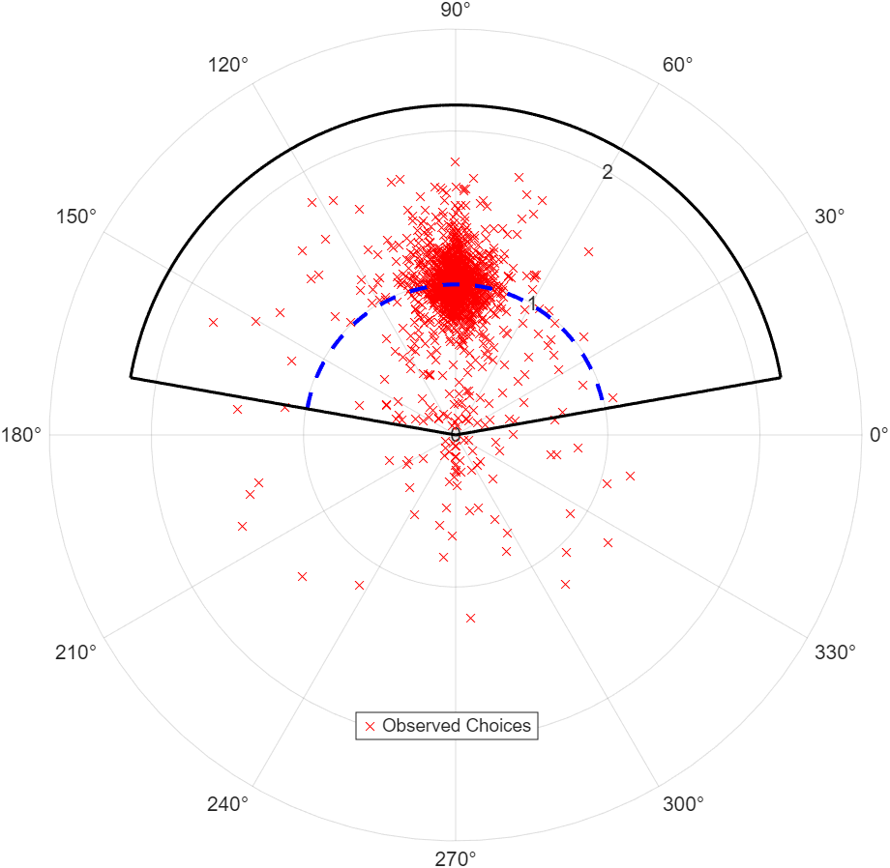}
    \caption{Distribution of all observed movement choices across the spatial grid.}
    \label{fig:grid_with}
\end{figure}

Following the general grid based approach introduced in \cite{alhaideri2025cyclistAAP}, 
the observed movement choices were derived from the step-to-step change in speed. For each pedestrian and at each 1s interval, we compute the relative change in speed increment using $\Delta s = \frac{d}{(s_{nt}\,\Delta t)}$, where $d$ is the displacement between consecutive 
positions, $s_{nt}$ is the pedestrian's current speed, and $\Delta t = 1\,\text{s}$. 
This produces a non-negative measure that reflects how much the pedestrian updated their speed relative to their current motion. The spatial choice set structure is illustrated in Fig.~\ref{fig:grid_def}, where each point $P_{nt}$ denotes the position of pedestrian $n$ at time step $t$, and the feasible movement adjustments around $P_{nt}$ are organized in a 
two ring grid. Based on the empirical distribution shown in Fig.~\ref{fig:grid_with}, the speed change boundaries were defined as a lower ring spanning $[0,\,0.99]$ representing modest adjustments, and an upper ring spanning $[0.99,\,2.17]$ representing stronger accelerations.

Observed choices were then assigned directly from these rings. Values falling within the lower ring were coded as choice~1, and values in the upper ring were coded as choice~2. 
Steps where $\Delta s$ exceeded the empirical upper bound, or where the change in heading angle exceeded $\pm 80^\circ$, were treated as invalid and excluded. The final number of valid choices used in the model was 1,875, including 938 observations for choice~1 (deceleration) and 937 observations for choice~2 (acceleration).

To formalize the behavioural structure of the model, the ResLogit expresses the utility of alternative $i$ chosen by individual $n$ at time step $t$ as the sum of a systematic component and two sources of unobserved variation expressed as:
\begin{equation}
U_{int} = V_{int} + g_{int} + \varepsilon_{int}.
\end{equation}
where $V_{int}$ is the deterministic utility specified as a linear function of observed attributes, $g_{int}$ is the residual neural component that introduces nonlinear transformations, and $\varepsilon_{int}$ is an i.i.d.\ extreme value error term. 
Following the standard random utility framework, pedestrians are assumed to select the alternative with the highest perceived utility. The ResLogit model therefore computes choice probabilities by combining the linear behavioural component $V_{int}$ with the nonlinear residual term $g_{int}$, while $\varepsilon_{int}$ captures the remaining unexplained variability. Full mathematical details of the residual architecture, its layer structure, and its probabilistic interpretation can be found in \cite{wong2021reslogit}.

\section{Results and Discussions}

The ResLogit model was estimated using 1,875 pedestrian--AV interaction time steps, with 1,312 observations used for training and 563 for validation (70:30 split). Model performance is presented in Tables~\ref{tab:classification} and~\ref{tab:cmatrix}. 
The overall accuracy of 0.58 suggests that the model may capture a moderate portion of the variability in pedestrian speeding choices. 
This accuracy likely reflects the high variability in pedestrian micromovements and the absence of latent psychological cues. The model precision and recall appear reasonably balanced across the two classes. This indicates that the model does not seem to favour acceleration or deceleration disproportionately. The confusion matrix further suggests that both outcomes are predicted with similar levels of correctness, although a noticeable number of misclassifications still appear. 
A closer look at these errors may suggest several contributing factors. Many movement adjustments lie close to the empirical decision boundaries, making them difficult to classify, especially under small trajectory noise or heading fluctuations. The model also relies solely on observable kinematic cues. Unmeasured perceptual factors such as gaze, attention, or intent may influence behaviour. Compressing a continuous range of adjustments into a binary choice could mask intermediate cases. These elements may help explain the misclassifications, and future improvements may require additional perceptual or contextual features or a richer choice structure.

To assess numerical robustness, the ResLogit model is also estimated over 1,000 Monte--Carlo runs, and the resulting coefficient distributions are generally stable. This implies that the model behaves reliably under repeated stochastic training. Several additional variables are tested during model specification. In particular, the Euclidean distance between the pedestrian and the AV was included in multiple preliminary formulations, but it consistently produced counter-intuitive sign and remained statistically insignificant. A variable counting the number of other road users within 30m of the pedestrian was also examined. Although it was intended to capture local traffic density, its effect was weak and unstable across specifications. This suggested limited behavioural value in the present dataset. Because these variables provided little interpretive insight and did not improve predictive performance, they were excluded from the final model.

Table~\ref{tab:reslogit_results} shows the estimated coefficients. Although all variables are statistically significant, the interpretation of their effects should remain tentative due to the inherent variability in the data and the modest predictive performance. 
The two relative speed terms have negative coefficients. The instantaneous relative speed may indicate that pedestrians could be more inclined to slow down when the AV is closing in rapidly. This could reflect a conservative adjustment in speed. The three sample smoothed relative speed term is even more negative. This might suggest that short term trends in approach speed influence behaviour more strongly than the instantaneous value alone. 
The visual looming rate shows a positive coefficient. This may imply that, when the AV appears to expand quickly in the pedestrian's field of view, pedestrians could be more inclined to accelerate, possibly to clear the vehicle's path sooner. 
However, this behaviour may also reflect reaction patterns specific to midblock encounters in constrained environments, and further work would be needed to generalize this tendency. The remaining distance to the destination also has a positive coefficient. This implies that pedestrians who are still far from completing the crossing may respond by accelerating rather than slowing down. This may reflect a desire to reduce exposure time.

The frontal and rear CRP indicators show a clear directional asymmetry. The positive coefficient for frontal CRP could indicate that pedestrians sometimes respond to a frontal threat by speeding up rather than slowing down. Conversely, the negative coefficient for rear CRP may imply that threats from behind encourage deceleration. This difference between forward and rear facing interactions might be related to how pedestrians perceive and react to approaching objects in their field of view. 
The lagged CRP terms add another layer of potential temporal influence. The three second lag of frontal CRP is negative, which contrasts with the positive instantaneous effect. This discrepancy may indicate that while immediate frontal risk encourages a rapid movement adjustment, sustained exposure leads to more cautious behaviour. A similar, though weaker, pattern appears in the lagged rear CRP term. These temporal dynamics could suggest that pedestrians integrate short term history into their decisions. Of course, all of these patterns could be context dependent, and it is difficult to draw broader behavioural conclusions from a single dataset.

\begin{table}[!t]
\centering
\caption{Performance metrics of ResLogit.}
\label{tab:classification}
\begin{tabular}{lcccc}
\toprule
\textbf{Class} & \textbf{Precision} & \textbf{Recall} & \textbf{F1-score} & \textbf{Support} \\
\midrule
Decelerate (0) & 0.60 & 0.57 & 0.58 & 293 \\
Accelerate (1) & 0.56 & 0.59 & 0.57 & 270 \\
\textbf{Accuracy} & \multicolumn{4}{c}{0.58 (N = 563)} \\
\bottomrule
\end{tabular}
\end{table}

\begin{table}[!t]
\centering
\caption{Confusion matrix for ResLogit.}
\label{tab:cmatrix}
\begin{tabular}{lcc}
\toprule
\textbf{True / Predicted} & \textbf{Decelerate} & \textbf{Accelerate} \\
\midrule
Decelerate & 167 & 126 \\
Accelerate & 112 & 158 \\
\bottomrule
\end{tabular}
\end{table}

\begin{table}[!t]
\centering
\caption{Estimated coefficients of ResLogit.}
\label{tab:reslogit_results}
\begin{tabular}{lccc}
\toprule
\textbf{Variable} & \textbf{Coefficient} & \textbf{Std.\ Error} & \textbf{\textit{t}-stat} \\
\midrule
Relative Speed Ped--AV & -0.187 & 0.0121 & -15.45 \\
Relative Speed Ped--AV (3s lag) & -0.661 & 0.0136 & -48.52 \\
Visual looming rate  & 1.297 & 0.0126 & 103.02 \\
Remaining distance to destination & 0.099 & 0.0139 & 7.08 \\
Front CRP & 1.900 & 0.0132 & 144.30 \\
Rear CRP & -3.044 & 0.0130 & -235.01 \\
Front CRP (3s lag) & -2.436 & 0.0141 & -172.97 \\
Rear CRP (3s lag) & -0.640 & 0.0138 & -46.23 \\
\bottomrule
\end{tabular}
\end{table}

\begin{figure}[!t]
\centering
\includegraphics[scale=0.55]{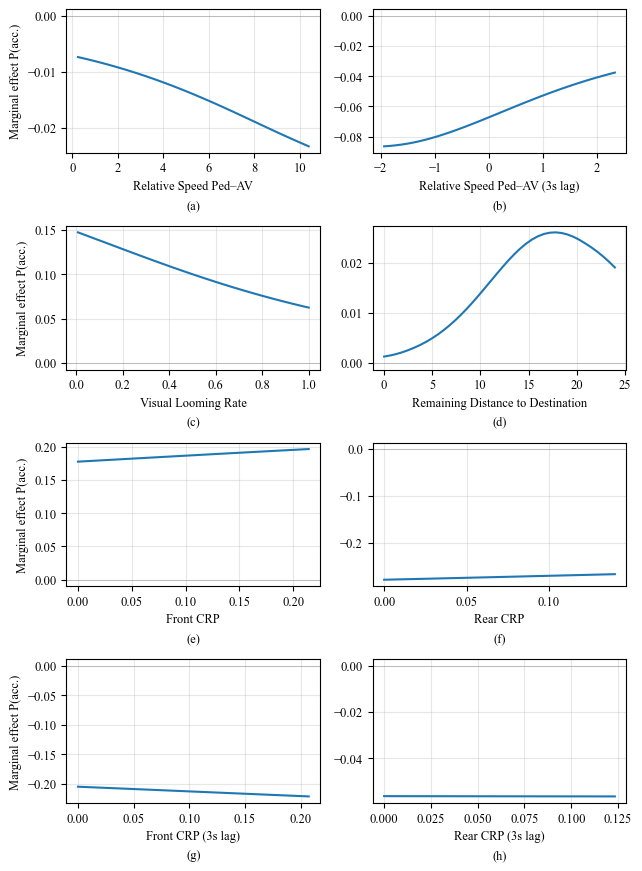}
\caption{Marginal effects showing how changes in each variable influence the probability of accelerating.}
\label{fig:marginal}
\end{figure}

\begin{figure}[!t]
\centering
\includegraphics[scale=0.55]{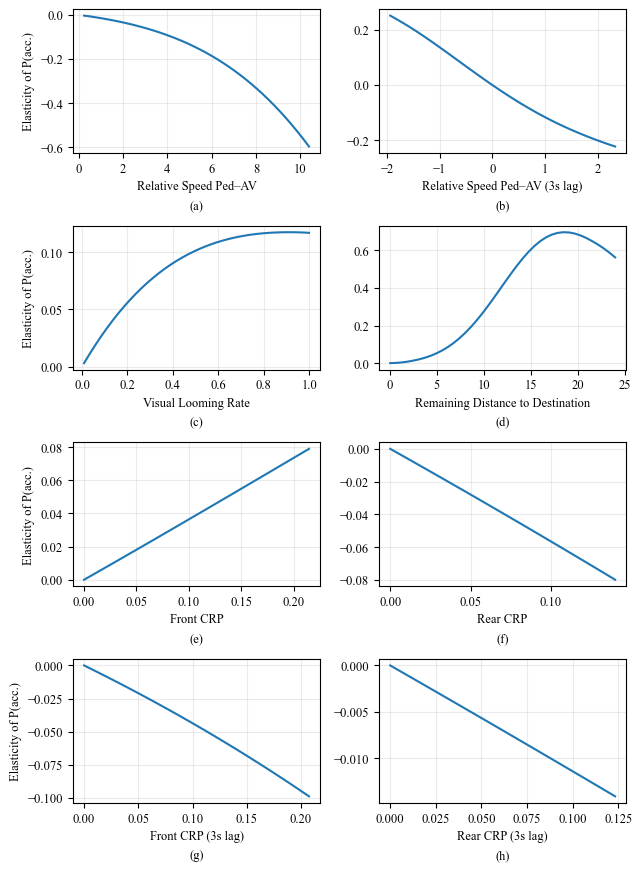}
\caption{Elasticities showing the percentage change in the probability of accelerating for a 1\% change in each explanatory variable.}
\label{fig:elast}
\end{figure}

Marginal effects, shown in Fig.~\ref{fig:marginal}, provide an estimate of how small changes in each explanatory variable may influence the probability of choosing acceleration. These effects complement the coefficient signs in Table~\ref{tab:reslogit_results} by illustrating how behavioural sensitivity varies across each variable. For both relative speed measures, the marginal effects are consistently negative, matching the coefficient signs. Higher instantaneous or recent relative speed appears to reduce the likelihood of accelerating, with the stronger lagged effect aligning with its larger negative coefficient. 
The visual looming rate shows positive marginal effects, consistent with its positive coefficient. The influence appears stronger at lower looming levels and tapers as looming increases. For the remaining distance to destination, the marginal effects are positive but show a clear peak around 15--20\,m. This suggests a possible mid-crossing zone where acceleration becomes more likely, after which the effect tapers. This pattern is broadly consistent with the positive coefficient but indicates that the influence may not increase linearly with distance.
The CRP indicators show a clear directional asymmetry. Frontal CRP has positive marginal effects, whereas rear CRP is strongly negative, consistent with their respective coefficients and suggesting opposite tendencies under forward  versus rear facing risk. The lagged CRP terms follow a similar pattern. Lagged frontal CRP becomes negative, contrasting with the instantaneous effect, while lagged rear CRP remains negative but weaker. This may indicate that the influence of perceived risk diminishes over short temporal windows. Across all variables, the magnitude of the marginal effects varies noticeably. The strongest influences appear for the instantaneous CRP terms, with frontal CRP increasing acceleration probability and rear CRP reducing it. Visual looming and remaining distance also show meaningful effects, with remaining distance exhibiting a clear mid-crossing peak. The instantaneous relative speed term shows only small marginal shifts. The lagged relative speed effect is somewhat stronger but still modest relative to the CRP measures. In general, the directional conflict indicators appear to exert the most influence, while relative speed inputs contribute more limited adjustments.

Elasticities, shown in Fig.~\ref{fig:elast}, indicate the percentage change in the probability of accelerating for a one percent change in each variable. The marginal effects describe absolute sensitivity, and the elasticities provide a relative measure that accounts for the scale of each input. The elasticity patterns broadly reflect the marginal effects. Both relative speed measures show negative elasticities, and the instantaneous effect being modest and the lagged effect slightly stronger. The visual looming rate has positive elasticities that rise at lower values and plateau. This implies a diminishing influence at high looming levels. The remaining distance to destination produces the largest elasticities among all variables. This suggests that mid-crossing distance may be a particularly sensitive region for acceleration adjustments. The CRP elasticities again show directional asymmetry. The frontal CRP exhibits positive elasticities. The rear CRP remains negative across its range. The lagged CRP terms follow the same pattern but with smaller magnitudes, which indicates a weaker influence of past risk compared with instantaneous exposure.
These patterns tentatively suggest that pedestrians may not rely on a single source of information when adjusting their movement. Instead, the evidence points toward multiple behavioural tendencies operating in parallel. The opposing effects of frontal and rear CRP imply distinct reactions to risk depending on its spatial origin. This may reflect different cognitive appraisals of forward  versus backward facing threat. The non-linear influence of remaining distance further implies that pedestrians may shift strategy mid-crossing, perhaps balancing progress toward the destination with perceived vehicle risk. In contrast, the relatively marginal influence of relative speed inputs suggests that speed cues alone might play a secondary role in these moment-to-moment adjustments. We hypothesize that the data capture more than one behavioural traits, one driven largely by risk perception and another by movement efficiency.

Although these insights are derived from trajectory data and a hybrid DCM–machine learning model, they offer tentative implications for pedestrian–AV interaction. The directional asymmetry in the CRP measures suggests that pedestrians could evaluate forward  and rear facing risk differently. If confirmed, AV yielding or communication strategies may benefit from emphasizing cues aligned with forward facing risk perception rather than assuming symmetric responses. The non-linear influence of remaining distance also hints at a mid-crossing adjustment zone in which pedestrians shift from early progression to more risk aware behaviour. This has potential relevance for AV prediction models, which often assume smoother patterns across the entire crossing. 
The relatively modest role of relative speed cues, compared with looming and CRP indicators, may indicate that pedestrians prioritize risk salient information over purely kinematic differences. This, together with the combined effects of distance and risk, suggests that multiple behavioural tendencies may operate simultaneously, driven by safety perception and another by movement efficiency. Models capable of capturing such blended mechanisms could complement existing AV prediction frameworks. Finally, the contrast between instantaneous and lagged CRP effects implies that pedestrians may integrate short term exposure history rather than reacting only to current conditions, a pattern that future AV systems may need to consider when modelling pedestrian responses.

\section{Conclusions}

This study examined how pedestrians adjust their moment-to-moment movement when interacting with an approaching AV at midblock crossings. Using the ResLogit framework, we modelled instantaneous acceleration and deceleration choices based on spatial, temporal, and perceptual indicators extracted from real world trajectories. The model achieved moderate predictive accuracy, suggesting that while behavioural structure is present, considerable variability remains unexplained.
The estimated coefficients indicate that several factors may shape pedestrian speed regulation. This includes relative speed, visual looming, directional exposure to collision risk, and the remaining distance to the destination. Short term lagged variables also appear to contribute, though their interpretation requires caution.
The findings point to potentially rich behavioural patterns but should not be viewed as definitive given the model's modest performance and the inherent variability of pedestrian motion. Future work could incorporate additional perceptual cues, contextual characteristics such as demographics, traffic density, or neural network components that capture higher order temporal effects. Such extensions may help clarify how pedestrians integrate instantaneous cues, recent movement history, directional threats, and crossing progress when interacting with AVs.

The model presented in this paper does not explicitly incorporate the influence of other road users present in the scene. These include human driven cars, trucks, cyclists, motorcyclists, and other pedestrians that move simultaneously with the pedestrian and the AV. 
Their trajectories and kinematics should be included in future work, since each road user type may affect pedestrian movement in a different and meaningful way. 
In addition, the geometric characteristics of the road, such as lane widths, crossing angles, and available refuge space, were not considered and may play an important role in shaping movement adjustments. 
Group behaviour also remains unmodelled. Pedestrians often cross in pairs or clusters, and these social dynamics may alter speed regulation, risk perception, and trajectory choices. Environmental conditions such as time of day and weather were likewise not examined, even though lighting and surface conditions can influence both perception and movement responses.
Although the current model focuses on binary acceleration decisions, the framework could be extended to support a richer set of behavioural alternatives. 
These may include low, moderate, and high acceleration choices, multiple levels of deceleration, speed maintenance, or turning manoeuvres. Studying such expanded alternatives, together with more detailed perceptual inputs, multi-agent interactions, roadway geometry, and group-level dynamics, could provide a more comprehensive representation of how pedestrians adjust their movement in dynamic and complex traffic environments.

\bibliographystyle{IEEEtran}
\bibliography{IEEEfull}

\end{document}